\documentclass[english,journal=jpcbfk,manuscript=article]{achemso}
\usepackage[T1]{fontenc}
\usepackage[latin9]{inputenc}
\usepackage{amsmath}
\usepackage{graphicx}
\usepackage{esint}

\makeatletter

\title{Generic Schemes for Single-Molecule Kinetics. 2: Information Content
of the Poisson Indicator}
\author{Thomas Avila}
\altaffiliation{Contributed equally to this work}
\author{D. Evan Piephoff}
\altaffiliation{Contributed equally to this work}
\author{Jianshu Cao}
\affiliation{Department of Chemistry, Massachusetts Institute of Technology, Cambridge,
MA 02139, United States}
\email{jianshu@mit.edu}

\usepackage{breqn}

\makeatother

\usepackage{babel}
\begin{document}
\begin{abstract}
Recently, we described a pathway analysis technique (paper 1) for
analyzing generic schemes for single-molecule kinetics based upon
the first-passage time distribution. Here, we employ this method to
derive expressions for the Poisson indicator, a measure of stochastic
variation (essentially equivalent to the Fano factor and Mandel's
Q parameter), for various renewal (memoryless) enzymatic reactions.
We examine its dependence on substrate concentration, without assuming
all steps follow Poissonian kinetics. Based upon fitting to the functional
forms of the first two waiting time moments, we show that, to second
order, the non-Poissonian kinetics are generally underdetermined but
can be specified in certain scenarios. For an enzymatic reaction with
an arbitrary intermediate topology, we identify a generic minimum
of the Poisson indicator as a function of substrate concentration,
which can be used to tune substrate concentration to the stochastic
fluctuations and estimate the largest number of underlying consecutive
links in a turnover cycle. We identify a local maximum of the Poisson
indicator (with respect to substrate concentration) for a renewal
process as a signature of competitive binding, either between a substrate
and an inhibitor or between multiple substrates. Our analysis explores
the rich connections between Poisson indicator measurements and microscopic
kinetic mechanisms.
\end{abstract}

\section{I. Introduction}

Single-molecule spectroscopy techniques have allowed the study of
single biomolecular complexes at a level of detail previously unattainable.\cite{Park2007}
Escaping the averaging of measured quantities inherent in ensemble
measurements, single-molecule studies offer insights into the details
of the dynamic behavior of biomolecules.\cite{Chen_2003,Deindl_2012}
In particular, these studies provide information on the underlying
kinetic scheme that is unavailable through traditional, bulk measurements
of chemical kinetics.\cite{Shen_2014,Press__2013} At their core,
single-molecule studies of enzymes and motor proteins interrogate
the waiting time between reaction events, such as the conversion of
substrate to product or the stepping of a motor protein along a filament.
The waiting time varies stochastically over the course of the observation
of the molecule, and sufficiently long time traces allow the waiting
time probability distribution to be described.\cite{Cao2000} Given
a kinetic mechanism, a mathematical expression for this waiting time
distribution in terms of kinetic parameters, such as rate constants
and reactant or product concentrations, may be derived. Furthermore,
expressions for the moments of the distribution and the correlations
between events may be obtained and compared to experimental observations.

From a theoretical standpoint, it is important to first determine
the information content available from single-molecule data and then
make connections to a generic scheme. Previous work has addressed
the relation of single-molecule data to reaction network connectivity
and developed a mathematical framework for treating data within a
given reaction scheme.\cite{Cao2000,Qian_2002,Bruno2005,Flomenbom2005a,Gopich2006,Ge_2008,Kolomeisky_2011,Ochoa_2011}
In a complementary fashion, we have described a pathway analysis approach
to generic reaction schemes for single-molecule kinetics (paper 1).\cite{Cao2008}
In contrast to other approaches, pathway analysis may be easily adapted
to arbitrary reaction scheme topologies. This method provides a straightforward
prescription for decomposing a proposed scheme via two basic kinetic
motifs, sequential and branching. Secondly, our approach requires
no assumption of Poissonian kinetics (i.e., rate processes), allowing
each step to be treated with the greatest possible generality. As
in paper 1,\cite{Cao2008} the current study deals with renewal (memoryless)
processes and, as a result, does not capture memory effects in the
action of single enzymes, as described previously experimentally and
theoretically.\cite{Cao2000,English2005} A subsequent paper will
generalize our method to arbitrary nonrenewal processes (paper 3).

This previous work\cite{Cao2008} provided calculation of the first
waiting time moment (i.e., mean first-passage time) for generalized
enzymatic schemes, which is directly related to the turnover rate
for the process. The turnover rate and mean first-passage time can
be determined from ensemble-averaging; however, higher-order moments,
which contain information on the underlying kinetic scheme of the
enzymatic reaction,\cite{Kou2005,Jung2010,Yang2011,Chaudhury2013,Kumar_2015}
are unique to single-molecule measurements. In particular, the Poisson
indicator, a normalized measure of stochastic fluctuations,\cite{Cao2008}
captures deviation from Poissonian statistics, taking on a positive
value for bunching behavior, a negative value for anti-bunching behavior,
and vanishing for a Poisson process. The dependence of the Poisson
indicator on substrate concentration can then inform which steps adhere
to or violate Poissonian statistics. Moreover, the Poisson indicator
is essentially equivalent to other normalized measures of the variance,
including Mandel's Q parameter from photon statistics\cite{Mandel1979},
the randomness parameter from studies of molecular motors\cite{Svoboda1994},
and the Fano factor.\cite{Daniels2009}

This paper is organized as follows: in section II, we extend the previously
introduced pathway analysis to the calculation of the second moment
of the waiting time distribution. We examine a generic model of enzymatic
reactions that can generate all possible kinetic models with the same
basic topological connectivity and contains no assumptions upon the
form of the kinetic scheme. As stated earlier, the only constraint
is that the overall reaction be a renewal process. In section III,
we employ this approach for the generic enzymatic reaction to evaluate
the maximal information content of measurements of the second waiting
time moment and, in particular, to examine the dependence of the second
moment on substrate concentration. Our results include functional
forms for the dependence of both the first (related to the turnover
rate) and second (related to the Poisson indicator) reaction waiting
time moments on substrate concentration, as well as explicit expressions
in terms of the waiting time moments for individual steps. We analyze
these functional forms and explore their connections to important
experimental limits. In section IV, we extend earlier, similar results\cite{Yang2011,Chaudhury2013}
to the more complex cases of competitive inhibition and competition
between multiple substrates. To our knowledge, these are the first
calculations of higher-order waiting time moments for these more complex
cases without assuming all steps follow Poissonian kinetics, and the
resulting expressions for the Poisson indicator differ qualitatively
from earlier results. In section V, we conclude.

\section{II. Self-consistent Pathway Analysis}

\begin{figure}[h]
\centering\includegraphics[width=3.25in]{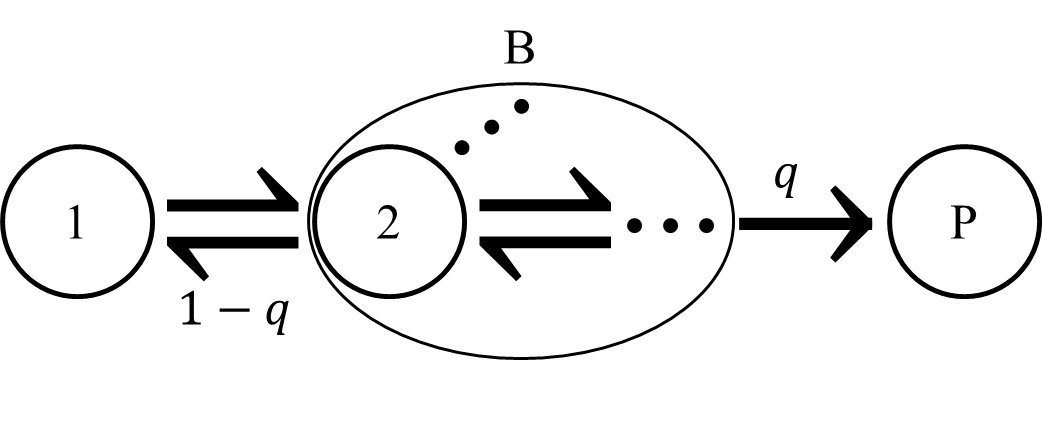}

\caption{\label{fig:GenChainRxn-1}Generic enzymatic reaction scheme. The aggregate
of intermediate states between the initial free enzyme (state 1) and
final transition(s) to product P, referred to as the bound/intermediate
state B, has an arbitrary internal topology. $q$ denotes the branching
probability for advancing to product from state 2.}
\end{figure}

Let $\phi(t)$ represent the waiting time distribution, which describes
the distribution of times between successive reaction events. The
moments of the waiting time distribution are given by $\left\langle t^{n}\right\rangle =\int_{0}^{\infty}t^{n}\phi(t)\,dt=(-1)^{n}\left.\frac{d^{n}\hat{\phi}(s)}{ds^{n}}\right|_{s=0}$,
where $\hat{\phi}(s)$ denotes the Laplace transform of $\phi(t)$,
defined as $\hat{\phi}(s)=\int_{0}^{\infty}e^{-st}\phi(t)\,dt$. Our
challenge is then to formulate the waiting time distribution for a
generic enzymatic reaction. The model we treat is illustrated in Figure
\ref{fig:GenChainRxn-1}. Here, states 1 and 2 are connected by a
reversible step, with an arbitrary topology after state 2, before
a final, irreversible transition (or set of transitions) to product
P. Upon the creation of a product molecule, we assume that the enzyme
regenerates quickly and irreversibly to state 1 (the initial free
enzyme state), where it begins another turnover. In our model, enzyme
turnover is a renewal process because it always begins in the same
state. In keeping with the Michaelis-Menten model of enzymatic reactions,
the first step corresponds to substrate binding to the enzyme,\cite{Cao2008}
which we assume to have a single substrate-binding site, making this
the only step with dependence on substrate concentration.
There may exist many intermediate underlying states between the substrate
binding step and final transition(s) to product. We refer to this
(possible) aggregate of states as the bound/intermediate state B,
which may undergo non-Poissonian decay due to its (possible) internal
dynamics, some of which may involve branching out of the chain as
well as cyclic loops.

Now, if we let $Q_{ij}(s)$ denote the waiting time distribution for
the $i$-to-$j$ transition in the Laplace domain, we can write the
overall waiting time (i.e., first-passage time) distribution in the
Laplace domain as
\begin{equation}
\hat{\phi}(s)=\frac{Q_{1P}(s)}{1-\tilde{Q}_{11}(s)}=Q_{1P}(s)\left[1+\tilde{Q}_{11}(s)+\tilde{Q}_{11}(s)^{2}+\tilde{Q}_{11}(s)^{3}+...\right]\label{eq:ChainRxnPDF-1}
\end{equation}
where $Q_{1P}(s)$ is the waiting time distribution for the passage
from state 1 to the product, and $\tilde{Q}_{11}(s)$ represents the
waiting time distribution for the passage out of and back to state
1.\cite{Cao2008} Each term in the infinite summation can be understood
as follows: the first term corresponds to the passage from state 1
to product P without returning to state 1, the second term corresponds
to the passage from state 1 to product P while returning to state
1 exactly once, the third term corresponds to the passage while returning
to state 1 exactly twice, and so on and so forth. Examining only the
initial free enzyme state and the bound/intermediate state, we can
write
\begin{align}
Q_{1P}(s) & =Q_{1B}(s)Q_{BP}(s)\\
\tilde{Q}_{11}(s) & =Q_{1B}(s)Q_{B1}(s)
\end{align}
where $Q_{1B}(s)$ is the waiting time distribution for substrate
binding, $Q_{B1}(s)$ is the waiting time distribution for substrate
unbinding, and $Q_{BP}(s)$ is the waiting time distribution for product
formation (i.e., the conversion of substrate to product after binding).
Now, the overall waiting time distribution is given by
\begin{equation}
\hat{\phi}(s)=\frac{Q_{1B}(s)Q_{BP}(s)}{1-Q_{1B}(s)Q_{B1}(s)}
\end{equation}
This scheme comprises a generic model for enzyme kinetics. It treats
explicitly the substrate binding step with waiting time distribution
$Q_{1B}(s)$, while treating in generality the decay of the bound/intermediate
state with the distributions $Q_{B1}(s)$ and $Q_{BP}(s)$.

In the Laplace domain, these waiting time distributions can be expanded
in terms of their moments as
\begin{equation}
Q_{ij}(s)=q_{ij}\left(1-s\left\langle t_{ij}\right\rangle +\frac{s^{2}}{2}\left\langle t_{ij}^{2}\right\rangle -\cdots\right)\label{eq:GenLaplaceDist}
\end{equation}
where the branching probabilities $q_{ij}$ account for the normalization
of probability, with $\sum_{j}q_{ij}=1$ and $\sum_{j}q_{ij}\left\langle t_{ij}^{k}\right\rangle =\left\langle \tau_{i}^{k}\right\rangle $,
the $k^{\textrm{th}}$ moment for the decay time of state $i$. Expanding
the overall waiting time distribution in terms of the moments for
the individual steps, as in eq \ref{eq:GenLaplaceDist}, yields
\begin{equation}
\hat{\phi}(s)=\frac{\alpha}{1-\beta}\label{eq:first_passage_time_distr}
\end{equation}
with
\begin{align}
\alpha & =q\left[1-s\left\langle \tau_{1}+t_{BP}\right\rangle +\frac{s^{2}}{2}\left\langle \left(\tau_{1}+t_{BP}\right)^{2}\right\rangle -\ldots\right]\\
\beta & =(1-q)\left[1-s\left\langle \tau_{1}+t_{B1}\right\rangle +\frac{s^{2}}{2}\left\langle \left(\tau_{1}+t_{B1}\right)^{2}\right\rangle -\ldots\right]
\end{align}
where $\left\langle \tau_{1}\right\rangle $ and $\left\langle \tau_{1}^{2}\right\rangle $
are the first and second waiting time moments for the decay of the
initial free enzyme state, $\left\langle t_{BP}\right\rangle $ and
$\left\langle t_{BP}^{2}\right\rangle $ are the first and second
waiting time moments for product formation, and $\left\langle t_{B1}\right\rangle $
and $\left\langle t_{B1}^{2}\right\rangle $ are the first and second
moments for substrate unbinding. The product formation branching probability
$q$ expresses the probability of advancing to product after substrate
binding. From an expression for the overall waiting time distribution,
the calculation of waiting time moments is straightforward. Given
$\hat{\phi}(s)=\frac{\alpha}{1-\beta}$, and observing that $\left.\left(1-\beta\right)\right|_{s=0}=\left.\alpha\right|_{s=0}$,
the first moment (i.e., mean first-passage time) is expressed as 
\begin{equation}
\left\langle t\right\rangle =-\left.\frac{d\hat{\phi}(s)}{ds}\right|_{s=0}=\left.\frac{-\left(\dot{\alpha}+\dot{\beta}\right)}{\alpha}\right|_{s=0}\label{eq:mean_first_pass_time_derv}
\end{equation}
where $\dot{x}$ denotes differentiation of $x$ with respect to the
Laplace variable.

While the mean first-passage time can be determined from bulk measurements,
higher-order moments, which contain information on microscopic mechanisms,\cite{Kou2005,Jung2010,Takagi2012a,Yang2011,Chaudhury2013,Kumar_2015}
are unique to single-molecule analysis. The Poisson indicator, which
measures stochastic fluctuations, is expressed as\cite{Cao2008} $\mathcal{Q}\left(t\right)=\frac{\left\langle N\left(t\right)^{2}\right\rangle -\left\langle N\left(t\right)\right\rangle ^{2}}{\left\langle N\left(t\right)\right\rangle }-1$,
where $\left\langle N\left(t\right)\right\rangle $ and $\left\langle N\left(t\right)^{2}\right\rangle $
are the first and second moments for the number of turnovers $N$
occurring within the measurement time window $t$. The first moment
$\left\langle N\left(t\right)\right\rangle $ is asymptotically related
to the mean first-passage time as\cite{Cox_1962} $\left\langle N\left(t\right)\right\rangle \sim t/\left\langle t\right\rangle $.
We are interested in the long-time limit $\mathcal{P}\equiv\lim_{t\rightarrow\infty}\mathcal{Q}\left(t\right)$,
which we simply refer to hereafter as the Poisson indicator (essentially
equivalent to the Fano factor\cite{Daniels2009} and Mandel's Q parameter\cite{Mandel1979}).
Asymptotically, $N\left(t\right)$ is Gaussian distributed for a renewal
process, with\cite{Cox_1962} $\left\langle N\left(t\right)^{2}\right\rangle -\left\langle N\left(t\right)\right\rangle ^{2}\sim\frac{\left\langle t^{2}\right\rangle -\left\langle t\right\rangle ^{2}}{\left\langle t\right\rangle ^{3}}t$,
resulting in\cite{Barkai_2004}

\begin{equation}
\mathcal{P}=\frac{\left\langle t^{2}\right\rangle -2\left\langle t\right\rangle ^{2}}{\left\langle t\right\rangle ^{2}}
\end{equation}
The Poisson indicator describes the deviation of a statistical process
from Poissonian behavior, assuming a positive value for the bunching
of events (super-Poissonian statistics), a negative value for the
anti-bunching of events (sub-Poissonian statistics), and vanishing
for a Poisson process. This quantity and equivalent measures of variation
are frequently calculated in experimental studies and can serve to
indicate the presence of dynamic disorder in particular reaction steps.\cite{Kou2005}
Of particular interest, the sign of the Poisson indicator yields information
about the topology of the kinetic mechanism: negative values of $\mathcal{P}$
correspond to kinetics dominated by sequential, multi-step reactions,
while positive values of $\mathcal{P}$ are associated with kinetics
dominated by a competing trapping process.\cite{Chaudhury2013,Moffitt2013}
In fact, when no branching occurs out of an enzymatic chain with an
irreversible final step, $\mathcal{P}\leq0$,\cite{Moffitt2013}.
Given the above functional form for $\hat{\phi}(s)$, the numerator
of the Poisson indicator can be calculated as
\begin{equation}
\left\langle t^{2}\right\rangle -2\left\langle t\right\rangle ^{2}=\left.\frac{\ddot{\alpha}+\ddot{\beta}}{\alpha}-\frac{2\dot{\alpha}}{\alpha^{2}}\left(\dot{\alpha}+\dot{\beta}\right)\right|_{s=0}\label{eq:Poiss_indic_num}
\end{equation}

\section{III. Generic Enzymatic Reaction}

\subsection*{A. Functional Forms and Parameter Specification}

Applying eqs \ref{eq:first_passage_time_distr}-\ref{eq:mean_first_pass_time_derv}
and \ref{eq:Poiss_indic_num} to the generic model of enzyme catalysis
(Figure \ref{fig:GenChainRxn-1}) yields
\begin{align}
\left\langle t\right\rangle  & =\frac{1}{q}\left[\left\langle \tau_{1}\right\rangle +\left\langle \tau_{B}\right\rangle \right]\\
\left\langle t^{2}\right\rangle -2\left\langle t\right\rangle ^{2} & =\frac{1}{q}\left[\left\langle \tau_{1}^{2}\right\rangle -2\left\langle \tau_{1}\right\rangle ^{2}+\left\langle \tau_{B}^{2}\right\rangle -2\left\langle t_{BP}\right\rangle \left(\left\langle \tau_{1}\right\rangle +\left\langle \tau_{B}\right\rangle \right)\right]
\end{align}
where $\left\langle \tau_{B}\right\rangle =q\left\langle t_{BP}\right\rangle +\left(1-q\right)\left\langle t_{B1}\right\rangle $
and $\left\langle \tau_{B}^{2}\right\rangle =q\left\langle t_{BP}^{2}\right\rangle +\left(1-q\right)\left\langle t_{B1}^{2}\right\rangle $
are the first and second waiting time moments, respectively, for bound/intermediate
state decay. In order to connect the above expressions to experimental
determinations of the Poisson indicator, we must examine their dependence
on substrate concentration $\left[\mathrm{S}\right]$. This dependence
can be addressed by treating substrate binding as a pseudo-first-order
rate step, which implies that substrate binding is a Poisson process
(i.e., $\left\langle \tau_{1}^{2}\right\rangle -2\left\langle \tau_{1}\right\rangle ^{2}=0$)
and $\left\langle \tau_{1}\right\rangle =\frac{1}{k_{1B}}$, with
pseudo-first-order rate $k_{1B}=k_{1B}^{\circ}\left[\textrm{S}\right]$,
where $k_{1B}^{\circ}$ is the rate constant for substrate binding.
Experimental studies of single enzymes have confirmed the validity
of this assumption,\cite{English2005} and its application leads to
\begin{align}
\left\langle t\right\rangle  & =\frac{1}{q}\left[\frac{1}{k_{1B}^{\circ}\left[\textrm{S}\right]}+\left\langle \tau_{B}\right\rangle \right]\label{eq:MM_firstmoment}\\
\left\langle t^{2}\right\rangle -2\left\langle t\right\rangle ^{2} & =\frac{1}{q}\left[\frac{-2\left\langle t_{BP}\right\rangle }{k_{1B}^{\circ}\left[\textrm{S}\right]}+\left\langle \tau_{B}^{2}\right\rangle -2\left\langle \tau_{B}\right\rangle \left\langle t_{BP}\right\rangle \right]
\end{align}
Finally, the Poisson indicator for the enzymatic reaction is given
by
\begin{equation}
\mathcal{P}\left(\left[\mathrm{S}\right]\right)=\frac{q\left[\frac{-2k_{1B}^{\circ}\left\langle t_{BP}\right\rangle }{\left[\textrm{S}\right]}+\left(k_{1B}^{\circ}\right)^{2}\left(\left\langle \tau_{B}^{2}\right\rangle -2\left\langle \tau_{B}\right\rangle \left\langle t_{BP}\right\rangle \right)\right]}{\left(\frac{1}{\left[\textrm{S}\right]}+k_{1B}^{\circ}\left\langle \tau_{B}\right\rangle \right)^{2}}\label{eq:GenChnRxnPoissonInd}
\end{equation}
This result gives a general functional form for the substrate dependence
of the Poisson indicator under the assumption of pseudo-first-order
kinetics for substrate binding:
\begin{equation}
\mathcal{P}\left(\left[\mathrm{S}\right]\right)=\frac{\frac{A}{\left[\textrm{S}\right]}+B}{\left(\frac{1}{\left[\textrm{S}\right]}+C\right)^{2}}\label{eq:FcnFormPoissInd}
\end{equation}
for constants $A$, $B$, and $C$ independent of $\left[\textrm{S}\right]$,
with expressions
\begin{align}
A & =-2qk_{1B}^{\circ}\left\langle t_{BP}\right\rangle \\
B & =q\left(k_{1B}^{\circ}\right)^{2}\left(\left\langle \tau_{B}^{2}\right\rangle -2\left\langle \tau_{B}\right\rangle \left\langle t_{BP}\right\rangle \right)\label{eq:GenB}\\
C & =k_{1B}^{\circ}\left\langle \tau_{B}\right\rangle \label{eq:GenC}
\end{align}
This result is analogous to those reported elsewhere.\cite{Yang2011,Chaudhury2013,Chaudhury2014}

\begin{figure}[H]
\centering\includegraphics[width=3.25in]{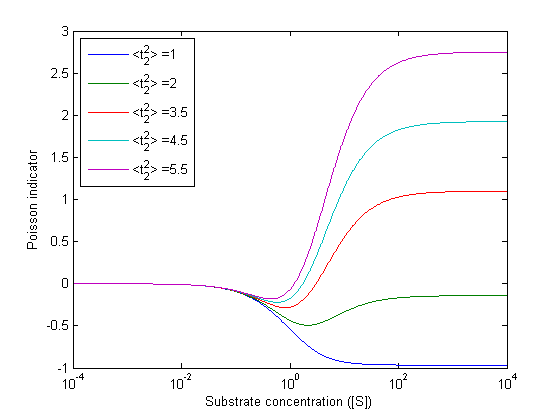}

\caption{\label{fig:BasicMM_Poissonplot}Plot of the Poisson indicator versus
substrate concentration for the generic enzymatic reaction (eq \ref{eq:GenChnRxnPoissonInd}).
The kinetic parameters chosen are $k_{1B}^{\circ}=1$, $q=0.5$, $\left\langle \tau_{B}\right\rangle =0.55$,
and $\left\langle t_{BP}\right\rangle =1$, with $2\left\langle \tau_{B}^{2}\right\rangle -0.03$
(represented by $\left\langle t_{2}^{2}\right\rangle $) given in
the legend.}
\end{figure}

From eqs \ref{eq:MM_firstmoment} and \ref{eq:GenChnRxnPoissonInd},
we see that, to second order, five parameters are needed to describe
the non-Poissonian kinetics of the generic enzymatic reaction (with
Poissonian binding): $k_{1B}^{\circ}$, $q$, $\left\langle \tau_{B}\right\rangle $,
$\left\langle t_{BP}\right\rangle $, and $\left\langle \tau_{B}^{2}\right\rangle $.
However, fitting measured data to these functional forms (with respect
to $\left[\textrm{S}\right]$) for the first waiting time moment and
Poisson indicator together only gives four independent parameters,
since $C$ (given in eq \ref{eq:GenC}) is not independent of the
two first moment parameters. Thus, to second order, the generic scheme
kinetics are underdetermined by one parameter. However, if $k_{1B}^{\circ}$
is known or can be estimated, then the kinetics can be specified.
Alternatively, if the enzyme is highly efficient (referred to as a
``perfectly evolved enzyme'' \cite{Wierenga2010}), such that the
turnover rate is limited only by the rate of diffusion of substrate
to the active site of the enzyme, we may assume that virtually every
substrate binding event leads to product. In our model, this corresponds
to $q\approx1$, which results in $\left\langle \tau_{B}\right\rangle \approx\left\langle t_{BP}\right\rangle $
and $\left\langle \tau_{B}^{2}\right\rangle \approx\left\langle t_{BP}^{2}\right\rangle $.
Now, three parameters are needed to describe the kinetics, and three
can be obtained from fitting (since $A$ is no longer independent
of the two first moment parameters); thus, the kinetics can be specified
under this assumption. Additionally, if the bound/intermediate state
undergoes Poissonian decay (i.e., the unbinding and product formation
transitions are rate steps), then $\left\langle \tau_{B}^{2}\right\rangle -2\left\langle \tau_{B}\right\rangle ^{2}=0$
and $\left\langle \tau_{B}\right\rangle =\left\langle t_{B1}\right\rangle =\left\langle t_{BP}\right\rangle $,
eliminating two kinetic parameters and causing $B$ (given in eq \ref{eq:GenB})
to vanish, thereby permitting the kinetics to be specified. It should
also be noted that our result for the first waiting time moment (eq
\ref{eq:MM_firstmoment}) follows the Michaelis-Menten functional
form; this is consistent with earlier work demonstrating that mechanisms
of arbitrary complexity yield a turnover rate with a hyperbolic dependence
on $\left[\textrm{S}\right]$ for zero conformational current.\cite{Cao2011,Wu2012}
Representative plots of the Poisson indicator versus $\left[\textrm{S}\right]$
appear in Figure \ref{fig:BasicMM_Poissonplot}. Qualitatively, the
Poisson indicator approaches finite limits at small and large $\left[\textrm{S}\right]$
and may feature a local minimum.

\subsection*{B. Minimum of Poisson Indicator and Topological Bound}

The Poisson indicator can be considered a measure of stochastic noise,
with sub-Poissonian statistics essentially corresponding to better
signal-to-noise than for a Poisson process. For nonzero $A$ and $C$,
when $B>0$, $\mathcal{P}\left(\left[\textrm{S}\right]\right)<0$
for $0<\left[\textrm{S}\right]<-A/B$; when $B\leq0$, $\mathcal{P}\left(\left[\textrm{S}\right]\right)<0$
for $\left[\textrm{S}\right]>0$. Thus, sub-Poissonian behavior is
achievable for any set of obtainable, nonzero $A$ and $C$, as there
always exists a (finite or infinite) range of substrate concentrations
at which stochastic fluctuations can enhance the statistics of generic
enzyme turnover (i.e., $\mathcal{P}<0$), even when branching occurs
within the bound/intermediate state. The Poisson indicator as a function
of $\left[\textrm{S}\right]$ has one stationary point at
\begin{equation}
\left[\mathrm{S}\right]^{*}=\left[k_{1B}^{\circ}\left(\frac{\left\langle \tau_{B}^{2}\right\rangle }{\left\langle t_{BP}\right\rangle }-\left\langle \tau_{B}\right\rangle \right)\right]^{-1}\label{eq:crit_S}
\end{equation}
which is only realizable when $\left[\textrm{S}\right]^{*}\geq0$.
For (i) $\left\langle \tau_{B}^{2}\right\rangle >\left\langle \tau_{B}\right\rangle \left\langle t_{BP}\right\rangle $,
$\mathcal{P}\left(\left[\textrm{S}\right]\right)$ is minimized at
$\left[\textrm{S}\right]^{*}$, where $\mathcal{P}\left(\left[\textrm{S}\right]^{*}\right)=-q\left\langle t_{BP}\right\rangle ^{2}/\left\langle \tau_{B}^{2}\right\rangle $,
which can never correspond to a local maximum. For (ii) $\left\langle \tau_{B}^{2}\right\rangle \leq\left\langle \tau_{B}\right\rangle \left\langle t_{BP}\right\rangle $,
$\mathcal{P}\left(\left[\textrm{S}\right]\right)$ is monotonic and
achieves a minimum of $q\left(\left\langle \tau_{B}^{2}\right\rangle -2\left\langle \tau_{B}\right\rangle \left\langle t_{BP}\right\rangle \right)/\left\langle \tau_{B}\right\rangle ^{2}$
as $\left[\textrm{S}\right]\rightarrow\infty$ (blue curve in Figure
\ref{fig:BasicMM_Poissonplot}). In either case, the minimum of $\mathcal{P}\left(\left[\textrm{S}\right]\right)$
essentially corresponds to the point of optimal signal-to-noise (with
respect to $\left[\textrm{S}\right]$); thus, $\left[\textrm{S}\right]$
can be tuned to the stochastic fluctuations to optimize enzyme turnover
statistics.

For the reaction of an enzyme with a single binding site and an irreversible
final step (or set of steps), the Poisson indicator is bounded by\cite{Barato_2015}
$\mathcal{P}\geq M_{\mathrm{max}}^{-1}-1$, where $M_{\mathrm{max}}$
is the largest value of $M$, the number of consecutive links in a
turnover cycle, with a network possibly containing multiple turnover
cycles. In our model, for a unicyclic network (which may still involve
branching within the bound/intermediate state), $M$ (and hence, $M_{\mathrm{max}}$)
corresponds to the number of underlying sequential (unbranched) rate
steps in the scheme; however, since the bound/intermediate state can
contain cyclic loops, and since multiple underlying transitions to
product can be present, the generic scheme can represent a multicyclic
network. The corresponding bound for $M_{\mathrm{max}}$ is given
by $M_{\mathrm{max}}\geq\left[\mathcal{P}+1\right]^{-1}$, which is
saturated when all links in the turnover cycle that corresponds to
$M_{\mathrm{max}}$ are irreversible with identical rates and the
rates of any branching steps out of this cycle are zero, which corresponds
to the longest homogeneous, sequential chain that can be formed in
the network.\cite{Barato_2015} This topological bound can be modified
using the minimum of $\mathcal{P}\left(\left[\textrm{S}\right]\right)$.
For (i) $\left\langle \tau_{B}^{2}\right\rangle >\left\langle \tau_{B}\right\rangle \left\langle t_{BP}\right\rangle $,
$M_{\mathrm{max}}$ is bounded by

\begin{equation}
M_{\mathrm{max}}\geq\left[1-q\frac{\left\langle t_{BP}\right\rangle ^{2}}{\left\langle \tau_{B}^{2}\right\rangle }\right]^{-1}\label{eq:N2boundPmin}
\end{equation}
 For (ii) $\left\langle \tau_{B}^{2}\right\rangle \leq\left\langle \tau_{B}\right\rangle \left\langle t_{BP}\right\rangle $,
we have 
\begin{equation}
M_{\mathrm{max}}\geq\left[1+q\frac{\left\langle \tau_{B}^{2}\right\rangle -2\left\langle \tau_{B}\right\rangle \left\langle t_{BP}\right\rangle }{\left\langle \tau_{B}\right\rangle ^{2}}\right]^{-1}\label{eq:N2boundPlim}
\end{equation}
Thus, eq \ref{eq:N2boundPmin} or \ref{eq:N2boundPlim} can be used
to estimate the largest number of underlying consecutive rate steps
in a turnover cycle. Notably, both of these bounds are independent
of $k_{1B}^{\circ}$, as are the inequalities identifying the two
cases. We note that even though the generic scheme kinetics are generally
underdetermined by one parameter, the expressions in eqs $\ref{eq:crit_S}$-\ref{eq:N2boundPlim},
along with the minimum of $\mathcal{P}\left(\left[\textrm{S}\right]\right)$
(and $\left\langle \tau_{B}^{2}\right\rangle /\left(\left\langle \tau_{B}\right\rangle \left\langle t_{BP}\right\rangle \right)$
to identify the case), can be evaluated from measurement of the first
two waiting time moments, without the need for any assumptions.

\subsection*{C. Limiting Behavior of Poisson Indicator}

The pathway analysis described above offers a simple route to the
calculation of waiting time moments, without the assumption of a particular
rate model. Ultimately, the goal is to connect experimental measurements
of waiting time moments to features of the underlying mechanism. From
the analytical expressions for the Poisson indicator as a function
of substrate concentration, we can now examine the experimentally
accessible limits.

\begin{figure}
\centering\textbf{(a)}

\textbf{\includegraphics[width=3.25in]{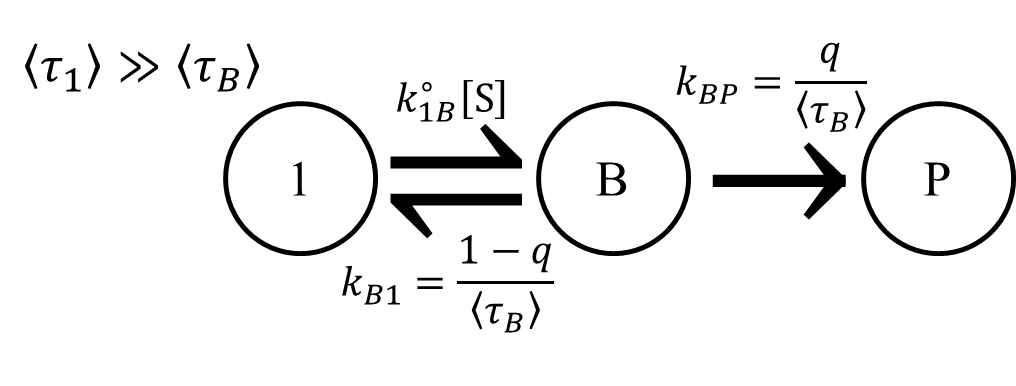}}

\textbf{(b)}

\includegraphics[width=3.25in]{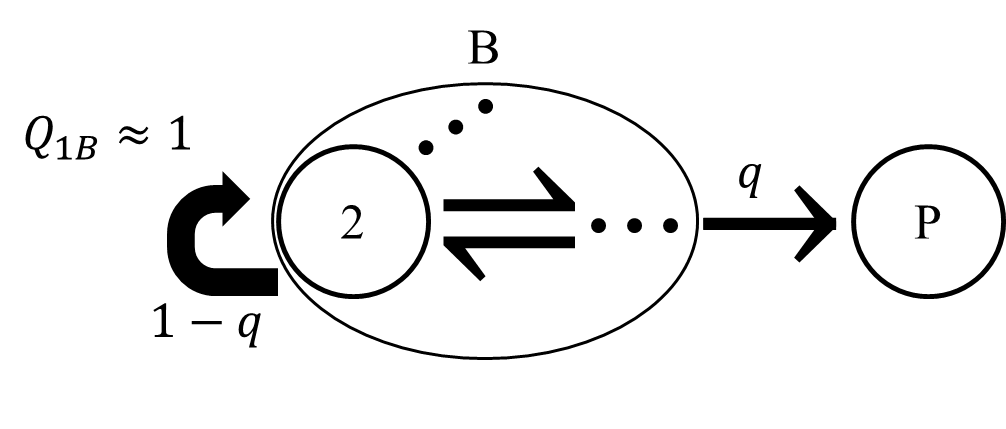}

\caption{\label{fig:high_low_S_schemes}Reduced representations of the generic
enzymatic scheme at low {[}(a){]} and high {[}(b){]} substrate concentration.
(a) At low substrate concentration, substrate binding is much slower
than bound/intermediate state decay ($\left\langle \tau_{1}\right\rangle \gg\left\langle \tau_{B}\right\rangle $),
resulting in the latter process becoming effectively Poissonian, i.e.,
the unbinding and product formation transitions behave as rate steps,
as the scheme reduces to a Michaelis-Menten model. (b) At high substrate
concentration, substrate binding effectively occurs instantaneously
($Q_{1B}\approx1$), as turnover begins in state 2 and unbinding proceeds
directly back into state 2.}
\end{figure}

As can be seen from eq \ref{eq:GenChnRxnPoissonInd}, in the limit
of low substrate concentration, the Poisson indicator vanishes. This
is a consequence of the assumption that substrate binding is a pseudo-first-order
rate process. At very low substrate concentration, substrate binding
becomes the rate determining step for the enzymatic process. Since
the Poisson indicator reflects the statistical properties of the waiting
time for the overall reaction, if the waiting time for the reaction
is dominated by a single step, the Poisson indicator will reflect
the statistical properties of that step. Hence, at very low substrate
concentration, the Poisson indicator vanishes. This is supported by
experimental observation of Poissonian kinetics for single enzymes
at very low substrate concentrations.\cite{English2005} As is also
apparent from eq \ref{eq:GenChnRxnPoissonInd}, at low substrate concentration,
we have, to leading order,
\begin{equation}
\mathcal{P}\left(\left[\mathrm{S}\right]\right)\approx-2qk_{1B}^{\circ}\left[\mathrm{S}\right]\left\langle t_{BP}\right\rangle \label{eq:small_S_limit}
\end{equation}
indicating that sub-Poissonian behavior, as well as a linear dependence
of the Poisson indicator on $\left[\textrm{S}\right]$, is always
expected at sufficiently low substrate concentration. This corresponds
to substrate binding being so much slower than bound/intermediate
state decay ($\left\langle \tau_{1}\right\rangle \gg\left\langle \tau_{B}\right\rangle $)
that the latter process becomes effectively Poissonian ($\left\langle \tau_{B}^{2}\right\rangle -2\left\langle \tau_{B}\right\rangle ^{2}\approx0$
and $\left\langle \tau_{B}\right\rangle \approx\left\langle t_{B1}\right\rangle \approx\left\langle t_{BP}\right\rangle $),
irrespective of the complexity of the underlying dynamics. That is,
the unbinding and product formation transitions behave as rate steps
with rates $k_{B1}=\left(1-q\right)/\left\langle \tau_{B}\right\rangle $
and $k_{BP}=q/\left\langle \tau_{B}\right\rangle $, respectively,
as the generic scheme reduces to the Michaelis-Menten scheme shown
in Figure \ref{fig:high_low_S_schemes}(a) (with $k_{1B}^{\circ}\left[\textrm{S}\right]\ll1/\left\langle \tau_{B}\right\rangle $),
resulting in sub-Poissonian statistics.

In the limit of high substrate concentration, the Poisson indicator
approaches a constant value. Substrate binding becomes arbitrarily
fast at high substrate concentrations, so the Poisson indicator will
reflect the statistical properties of the steps not dependent upon
substrate concentration. For the generic enzymatic reaction, the large-$\left[\textrm{S}\right]$
limit is given by 
\begin{equation}
\mathcal{P}_{\left[\textrm{S}\right]\rightarrow\infty}=q\frac{\left\langle \tau_{B}^{2}\right\rangle -2\left\langle \tau_{B}\right\rangle \left\langle t_{BP}\right\rangle }{\left\langle \tau_{B}\right\rangle ^{2}}\label{eq:largeSlimit}
\end{equation}
which is recovered when $\left\langle \tau_{1}\right\rangle \ll\left\langle \tau_{B}\right\rangle $.
This corresponds to instantaneous substrate binding ($Q_{1B}\approx1$),
with turnover effectively beginning in state 2 and unbinding proceeding
back into state 2, as shown in the reduced scheme in Figure \ref{fig:high_low_S_schemes}(b).
We note that $\mathcal{P}_{\left[\textrm{S}\right]\rightarrow\infty}$
vanishes when the bound/intermediate state is unaggregated (i.e.,
contains a single underlying state, undergoing Poissonian decay) and
can be positive when branching occurs within the bound/intermediate
state. The expression for $\mathcal{P}_{\left[\textrm{S}\right]\rightarrow\infty}$
can be simplified with basic assumptions about the nature of the enzymatic
system. Under the aforementioned perfectly evolved enzyme assumption
(in which $q\approx1$), the large-$\left[\textrm{S}\right]$ limit
of the Poisson indicator simplifies to
\begin{equation}
\mathcal{P}_{\left[\textrm{S}\right]\rightarrow\infty}\approx\frac{\left\langle t_{BP}^{2}\right\rangle -2\left\langle t_{BP}\right\rangle ^{2}}{\left\langle t_{BP}\right\rangle ^{2}}=\mathcal{P}_{BP}
\end{equation}
where we have defined $\mathcal{P}_{BP}$ as the Poisson indicator
for product formation. Therefore, for an enzyme of this type, determination
of the Poisson indicator at high substrate concentration directly
informs upon the statistical properties of the step(s) converting
substrate to product after substrate binding. In a similar vein, we
can consider the case of an enzyme where product formation is much
slower than substrate unbinding, which corresponds to the limit $q\rightarrow0$
in our model. The large-$\left[\textrm{S}\right]$ limit of the Poisson
indicator is then given by
\begin{align}
\mathcal{P}_{\left[\textrm{S}\right]\rightarrow\infty} & \approx\frac{q}{1-q}\frac{\left\langle t_{B1}^{2}\right\rangle -2\left\langle t_{B1}\right\rangle \left\langle t_{BP}\right\rangle }{\left\langle t_{B1}\right\rangle ^{2}}\nonumber \\
 & =\frac{q}{1-q}\left(\mathcal{P}_{B1}+\frac{2\left(\left\langle t_{B1}\right\rangle -\left\langle t_{BP}\right\rangle \right)}{\left\langle t_{B1}\right\rangle }\right)
\end{align}
where $\mathcal{P}_{B1}\equiv\frac{\left\langle t_{B1}^{2}\right\rangle -2\left\langle t_{B1}\right\rangle ^{2}}{\left\langle t_{B1}\right\rangle ^{2}}$.
Hence, in this case, the large-$\left[\textrm{S}\right]$ limit depends
upon the Poisson indicator for the substrate unbinding process and
a normalized measure of the difference in average waiting time for
substrate unbinding and product formation. These limits offer another
means of tying experimental measurements of the Poisson indicator
to the underlying statistics, in addition to the possibility of directly
fitting experimental data to the general functional form of the Poisson
indicator.

We now proceed to extend our approach to more complex reaction schemes.

\section{IV. Inhibition and Selective Binding}

\subsection*{A. Competitive Inhibition}

\begin{figure}[H]
\centering\includegraphics[width=3.25in]{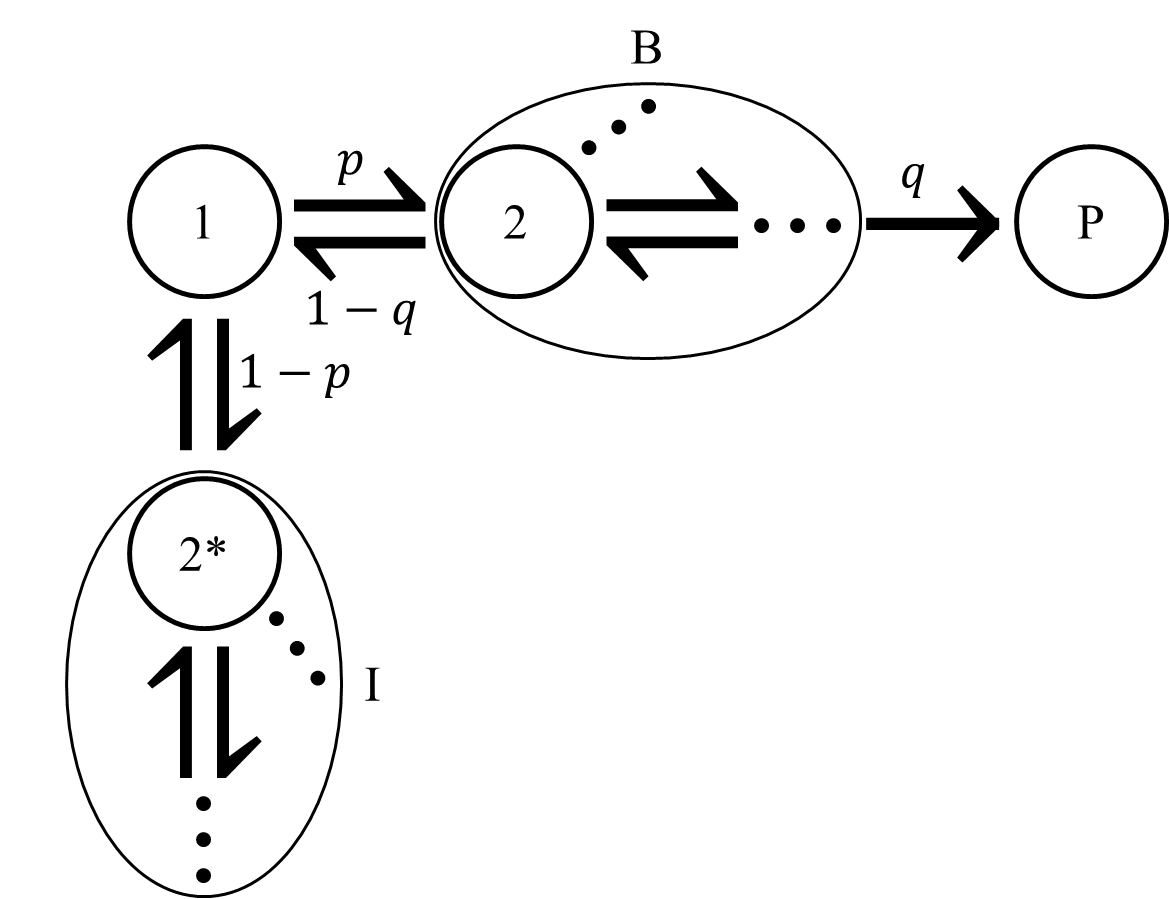}

\caption{\label{fig:GenInhibited}Generalized enzymatic reaction incorporating
competitively inhibited state I, which can be an aggregate of states
with an arbitrary internal topology. $p$ and $q$ are the branching
probabilities for binding substrate (versus inhibitor) and for advancing
to product from state 2, respectively.}
\end{figure}

As a further example of our approach, we examine a generalized scheme
for enzymatic reactions with competitive inhibition (Figure \ref{fig:GenInhibited}).
We note that inhibited single-molecule reactions have experimental
relevance\cite{Piwonski_2012} and have been the subject of theoretical
studies involving rate processes.\cite{Saha_2012,Chaudhury2014}
Now, the free enzyme may bind either substrate or inhibitor, reaching
state 2 or 2{*} with probability $p$ or $(1-p)$, respectively. Like
the bound/intermediate state, the inhibited state I may be an aggregate
of states with an arbitrary internal topology; thus, it may undergo
non-Poissonian decay. Following the same analysis as before, the overall
waiting time distribution takes the form 
\begin{equation}
\hat{\phi}(s)=\frac{Q_{1B}(s)Q_{BP}(s)}{1-Q_{1B}(s)Q_{B1}(s)-Q_{1I}(s)Q_{I1}(s)}=\frac{\alpha}{1-\beta}
\end{equation}
where $Q_{1B}(s)$ and $Q_{B1}(s)$ are the waiting time distributions
for substrate binding and unbinding, $Q_{1I}(s)$ and $Q_{I1}(s)$
are the distributions for inhibitor binding and unbinding, and $Q_{BP}(s)$
is the distribution for product formation. The constants $\alpha$
and $\beta$ are then
\begin{align}
\alpha & =pq\left[1-s\left\langle t_{1B}+t_{BP}\right\rangle +\frac{s^{2}}{2}\left\langle \left(t_{1B}+t_{BP}\right)^{2}\right\rangle -\ldots\right]\\
\beta & =p(1-q)\left[1-s\left\langle t_{1B}+t_{B1}\right\rangle +\frac{s^{2}}{2}\left\langle \left(t_{1B}+t_{B1}\right)^{2}\right\rangle -\ldots\right]+\nonumber \\
 & \,\,\,\,\,\,\,\,(1-p)\left[1-s\left\langle t_{1I}+\tau_{I}\right\rangle +\frac{s^{2}}{2}\left\langle \left(t_{1I}+\tau_{I}\right)^{2}\right\rangle -\ldots\right]
\end{align}
where $\left\langle t_{1B}\right\rangle $ and $\left\langle t_{1B}^{2}\right\rangle $
are the first and second moments for the substrate binding waiting
time, $\left\langle t_{1I}\right\rangle $ and $\left\langle t_{1I}^{2}\right\rangle $
are the first and second moments for the inhibitor binding waiting
time, and $\left\langle \tau_{I}\right\rangle $ and $\left\langle \tau_{I}^{2}\right\rangle $
are the first and second moments for the decay time of the inhibited
state. Again, we can examine the dependence on the concentrations
of substrate and inhibitor by assuming that the binding of each is
a rate process. This assumption leads to $\left\langle t_{1B}^{k}\right\rangle =\left\langle t_{1I}^{k}\right\rangle =\left\langle \tau_{1}^{k}\right\rangle $,
where $\left\langle \tau_{1}^{k}\right\rangle =p\left\langle t_{1B}^{k}\right\rangle +\left(1-p\right)\left\langle t_{1I}^{k}\right\rangle $,
with
\begin{align}
\left\langle \tau_{1}\right\rangle  & =\frac{1}{k_{1B}+k_{1I}}\label{eq:inhibitedt1}\\
p & =\frac{k_{1B}}{k_{1B}+k_{1I}}\label{eq:inhibitedp}
\end{align}
and $\left\langle \tau_{1}^{2}\right\rangle -2\left\langle \tau_{1}\right\rangle ^{2}=0$
for pseudo-first-order rate $k_{1I}=k_{1I}^{\circ}\left[\textrm{I}\right]$,
where $k_{1I}^{\circ}$ is the rate constant for inhibitor binding,
and $\left[\mathrm{I}\right]$ is the inhibitor concentration.

The first overall waiting time moment for the enzymatic reaction in
the presence of a competitive inhibitor is then
\begin{equation}
\left\langle t\right\rangle =\frac{1}{q}\left[\frac{1+k_{1I}^{\circ}\left[\mathrm{I}\right]\left\langle \tau_{I}\right\rangle }{k_{1B}^{\circ}\left[\textrm{S}\right]}+\left\langle \tau_{B}\right\rangle \right]\label{eq:compinhibfirstmom}
\end{equation}
Now, calculation of the Poisson indicator as before yields
\begin{equation}
\mathcal{P}\left(\left[\mathrm{S}\right]\right)=\frac{\frac{A}{\left[\textrm{S}\right]}+B}{\left(\frac{1}{\left[\textrm{S}\right]}+C\right)^{2}}\label{eq:InhFcnlForm}
\end{equation}
where $A$, $B$, and $C$ now depend upon the inhibitor concentration
and are given by
\begin{align}
A & =\frac{qk_{1B}^{\circ}}{\left(1+k_{1I}^{\circ}\left[\textrm{I}\right]\left\langle \tau_{I}\right\rangle \right)^{2}}\left[-2\left\langle t_{BP}\right\rangle +k_{1I}^{\circ}\left[\mathrm{I}\right]\left(\left\langle \tau_{I}^{2}\right\rangle -2\left\langle \tau_{I}\right\rangle \left\langle t_{BP}\right\rangle \right)\right]\label{eq:Inh_A}\\
B & =\frac{q\left(k_{1B}^{\circ}\right)^{2}}{\left(1+k_{1I}^{\circ}\left[\textrm{I}\right]\left\langle \tau_{I}\right\rangle \right)^{2}}\left[\left\langle \tau_{B}^{2}\right\rangle -2\left\langle \tau_{B}\right\rangle \left\langle t_{BP}\right\rangle \right]\label{eq:Inh_B}\\
C & =\frac{k_{1B}^{\circ}\left\langle \tau_{B}\right\rangle }{1+k_{1I}^{\circ}\left[\textrm{I}\right]\left\langle \tau_{I}\right\rangle }\label{eq:Inh_C}
\end{align}
Notably, this is the same basic functional form (with respect to $\left[\textrm{S}\right]$)
as that in the uninhibited case (eq \ref{eq:FcnFormPoissInd}). To
second order, eight parameters are needed to describe the non-Poissonian
kinetics (with Poissonian binding): $k_{1B}^{\circ}$, $k_{1I}^{\circ}$,
$q$, $\left\langle \tau_{B}\right\rangle $, $\left\langle \tau_{I}\right\rangle $,
$\left\langle t_{BP}\right\rangle $, $\left\langle \tau_{B}^{2}\right\rangle $,
and $\left\langle \tau_{I}^{2}\right\rangle $. However, eqs \ref{eq:compinhibfirstmom}-\ref{eq:Inh_C}
indicate that fitting (with respect to $\left[\textrm{S}\right]$
and $\left[\textrm{I}\right]$) to second order only gives six independent
parameters, making the kinetics underdetermined by two parameters.
However, the number of underdetermined parameters can be reduced in
several situations. (i) If either $k_{1B}^{\circ}$ or $k_{1I}^{\circ}$
is known, then one kinetic parameter can be eliminated (two if both
are known). (ii) If inhibitor unbinding is a rate process with rate
$k_{I1}$, then $\left\langle \tau_{I}\right\rangle =1/k_{I1}$ and
$\left\langle \tau_{I}^{2}\right\rangle -2\left\langle \tau_{I}\right\rangle ^{2}=0$,
which eliminates one kinetic parameter. (iii) If the aforementioned
perfectly evolved enzyme assumption holds, then the number of underdetermined
parameters is reduced by one (as shown in section III.A). (iv) If
the bound/intermediate state undergoes Poissonian decay, then the
number of underdetermined parameters is also reduced by one (as shown
in section III.A). Thus, the kinetics can be specified in a variety
of ways.

\begin{figure}
\centering \textbf{(a)}

\includegraphics[width=3.25in]{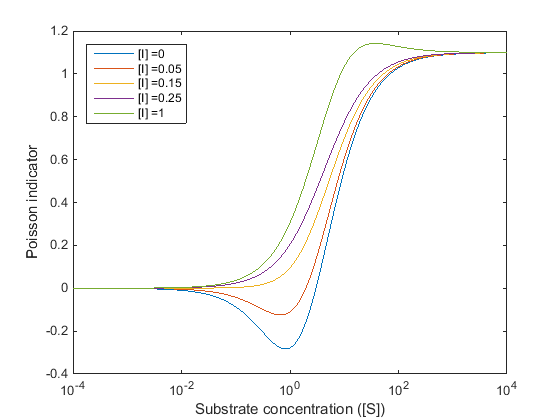}

\textbf{(b)}

\includegraphics[width=3.25in]{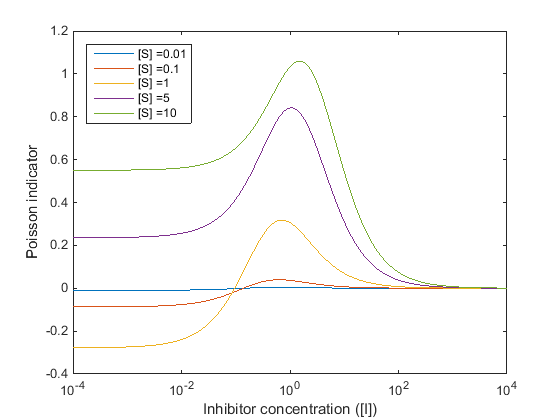}

\caption{\label{fig:GenEnzInhPlot}(a) Plot of the Poisson indicator versus
substrate concentration at fixed inhibitor concentration. The numerical
parameters are $k_{1B}^{\circ}=1$, $k_{1I}^{\circ}=1$, $q=0.5$,
$\left\langle \tau_{B}\right\rangle =0.55$, $\left\langle \tau_{I}\right\rangle =3$,
$\left\langle t_{BP}\right\rangle =1$, $\left\langle \tau_{B}^{2}\right\rangle =1.765$,
and $\left\langle \tau_{I}^{2}\right\rangle =20$, with $\left[\textrm{I}\right]$
given in the legend. (b) Plot of the Poisson indicator versus inhibitor
concentration at fixed substrate concentration. The numerical parameters
are identical to those in (a), except now $\left[\textrm{S}\right]$
is given in the legend.}
\end{figure}

Figure \ref{fig:GenEnzInhPlot}(a) illustrates the dependence of the
Poisson indicator on substrate concentration across a range of inhibitor
concentrations. As was the case for the uninhibited reaction, the
Poisson indicator vanishes at very low substrate concentration and
adopts the form given in eq \ref{eq:largeSlimit} at high substrate
concentration. The large-$\left[\textrm{S}\right]$ limits match for
these two cases because, from eqs \ref{eq:inhibitedt1} and \ref{eq:inhibitedp},
when $k_{1B}^{\circ}\left[\textrm{S}\right]\gg k_{1I}^{\circ}\left[\textrm{I}\right]$,
$\left\langle \tau_{1}\right\rangle \approx\frac{1}{k_{1B}^{\circ}\left[\textrm{S}\right]}$
and $p\approx1$. Qualitative differences are evident between Figures
\ref{fig:BasicMM_Poissonplot} and \ref{fig:GenEnzInhPlot}(a). In
particular, a local maximum (with respect to $\left[\textrm{S}\right]$)
can be achieved with a competitive inhibitor when $AC>B$ and $C>2B/A$.
This unique feature essentially corresponds to the point of poorest
signal-to-noise for a given, obtainable $A$, $B$, and $C$ (capable
of achieving one). We note that $\mathcal{P}\left(\left[\mathrm{S}\right]\right)$
may instead achieve a local minimum or no realizable local extremum.
In the presence of a competitive inhibitor, eq \ref{eq:small_S_limit}
for low $\left[\textrm{S}\right]$ does not generally apply (except
in the limit of vanishing $\left[\mathrm{I}\right]$). In fact, under
certain conditions, the Poisson indicator can be non-negative at all
substrate concentrations, precluding sub-Poissonian behavior. Similarly,
the Poisson indicator can be non-positive at all substrate concentrations
under certain conditions {[}behavior not shown in Figure \ref{fig:GenEnzInhPlot}(a){]},
even when a competitive inhibitor is present.

The inherent asymmetry between inhibitor and substrate is demonstrated
in Figure \ref{fig:GenEnzInhPlot}(b), where the Poisson indicator
is plotted against inhibitor concentration across a range of substrate
concentrations.
The Poisson indicator can attain a local maximum (with respect to
$\left[\textrm{I}\right]$), which corresponds to the point of optimal
inhibition (i.e., poorest signal-to-noise, essentially) for a given
set of conditions (under which one can be achieved). This extremum
is important in the context of drug design, since many drugs function
by acting as inhibitors. In such cases, $\left[\textrm{I}\right]$
can be selectively tuned to attain optimal inhibition statistics.
We note that $\mathcal{P}\left(\left[\mathrm{I}\right]\right)$ may
instead achieve a local minimum or no realizable local extremum {[}cases
not shown in Figure \ref{fig:GenEnzInhPlot}(b){]}. In the limit of
saturating $\left[\textrm{I}\right]$, $\mathcal{P}$ vanishes because
inhibitor binding becomes the only feasible transition. As is to be
expected, eqs \ref{eq:Inh_A}-\ref{eq:Inh_C} above reduce to the
results for the generic enzymatic reaction in the limit of vanishing
$\left[\textrm{I}\right]$.

\subsection*{B. Multiple Substrates}

\begin{figure}
\centering\includegraphics[width=3.25in]{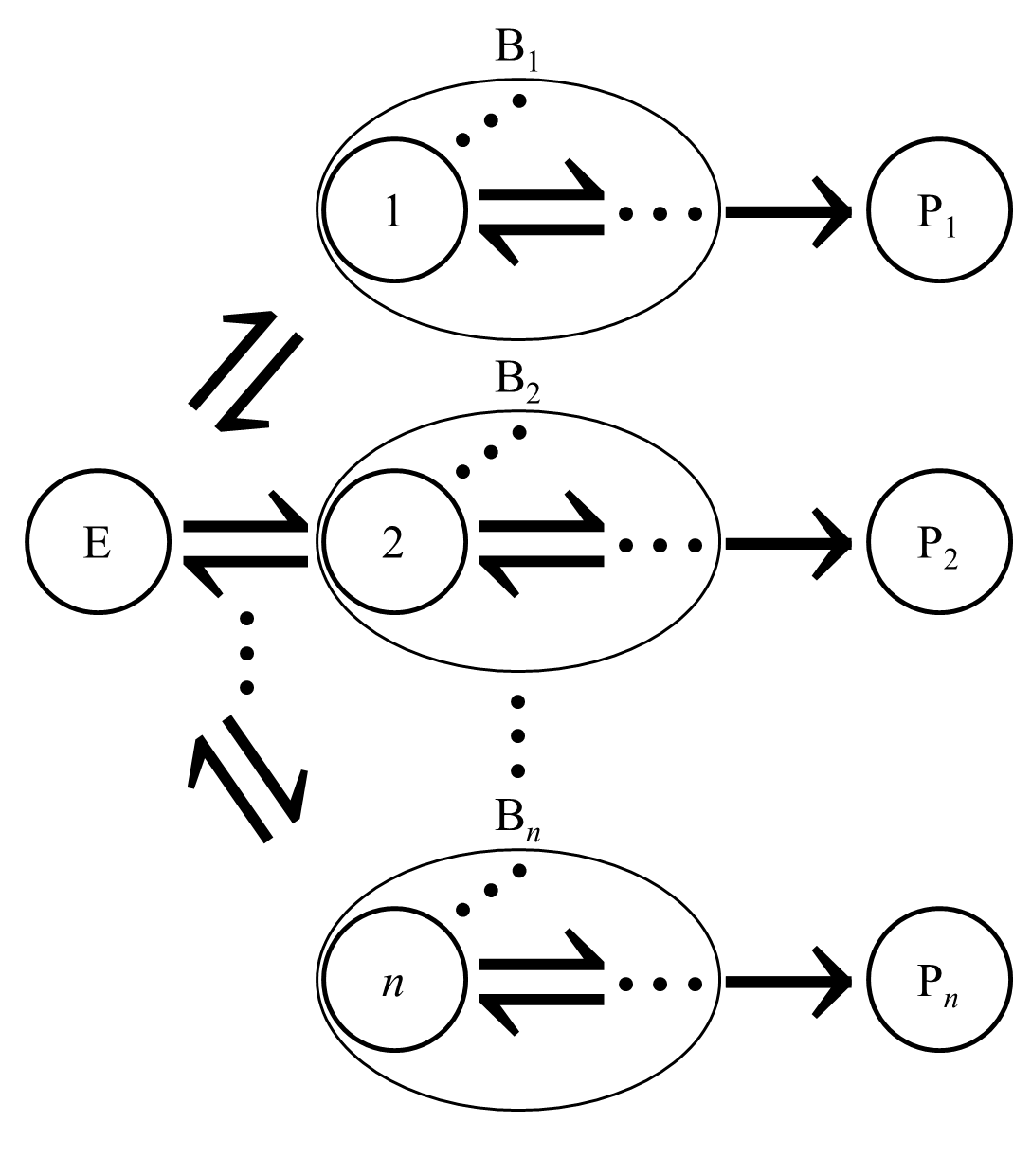}

\caption{\label{fig:MultPathways}Generalized enzymatic reaction featuring
$n$ competing substrates with concentrations $\left[\textrm{S}_{1}\right]$,
$\left[\textrm{S}_{2}\right]$, $\ldots$, $\left[\textrm{S}_{n}\right]$.}
\end{figure}

Our methodology can also be applied to more complex systems. In fact,
generalization to a reaction with multiple substrates is straightforward.
The scheme for this case is illustrated in Figure \ref{fig:MultPathways}.
The waiting time distribution for the conversion of any one of the
$n$ substrates to its corresponding product is given by
\begin{equation}
\hat{\phi}(s)=\frac{\sum_{i}Q_{EB_{i}}(s)Q_{B_{i}P}(s)}{1-\sum_{i}Q_{EB_{i}}(s)Q_{B_{i}E}(s)}=\frac{\alpha}{1-\beta}
\end{equation}
where $Q_{EB_{i}}(s)$ is the waiting time distribution for the binding
of substrate $\textrm{S}_{i}$, $Q_{B_{i}E}(s)$ is the distribution
for the unbinding of substrate $\textrm{S}_{i}$, and $Q_{B_{i}P}(s)$
is the distribution for the conversion of bound/intermediate state
$\mathrm{B}_{i}$ to the corresponding product $\mathrm{P}_{i}$.
In terms of the waiting time moments for the individual steps,
\begin{align}
\alpha & =\sum_{i}p_{i}q_{i}\left[1-s\left\langle t_{EB_{i}}+t_{B_{i}P}\right\rangle +\frac{s^{2}}{2}\left\langle \left(t_{EB_{i}}+t_{B_{i}P}\right)^{2}\right\rangle \right]\\
\beta & =\sum_{i}p_{i}\left(1-q_{i}\right)\left[1-s\left\langle t_{EB_{i}}+t_{B_{i}E}\right\rangle +\frac{s^{2}}{2}\left\langle \left(t_{EB_{i}}+t_{B_{i}E}\right)^{2}\right\rangle \right]
\end{align}
where $q_{i}$ is the branching probability for the formation of product
$\mathrm{P}_{i}$, $\left\langle t_{EB_{i}}\right\rangle $ and $\left\langle t_{EB_{i}}^{2}\right\rangle $
are the first and second waiting time moments for the binding of substrate
$\textrm{S}_{i}$, $\left\langle t_{B_{i}E}\right\rangle $ and $\left\langle t_{B_{i}E}^{2}\right\rangle $
are the first and second waiting time moments for the unbinding of
substrate $\textrm{S}_{i}$, and $\left\langle t_{B_{i}P}\right\rangle $
and $\left\langle t_{B_{i}P}^{2}\right\rangle $ are the first and
second moments for the formation of product $\mathrm{P}_{i}$. Assuming
that the binding of any substrate is a rate process, then $\left\langle t_{EB_{i}}\right\rangle =\left\langle \tau_{E}\right\rangle $
and $\left\langle t_{EB_{i}}^{2}\right\rangle =\left\langle \tau_{E}^{2}\right\rangle $,
where $\left\langle \tau_{E}\right\rangle =\sum_{i}p_{i}\left\langle t_{EB_{i}}\right\rangle $
and $\left\langle \tau_{E}^{2}\right\rangle =\sum_{i}p_{i}\left\langle t_{EB_{i}}^{2}\right\rangle $
are the first and second waiting time moments, respectively, for the
decay of the initial free enzyme state, with $p_{i}$ representing
the probability of binding substrate $\textrm{S}_{i}$. We now have
$p_{i}$ and $\left\langle \tau_{E}\right\rangle $ given by
\begin{align}
p_{i} & =\frac{k_{EB_{i}}}{\sum_{i}k_{EB_{i}}}\\
\left\langle \tau_{E}\right\rangle  & =\frac{1}{\sum_{i}k_{EB_{i}}}
\end{align}
as well as $\left\langle \tau_{E}^{2}\right\rangle -2\left\langle \tau_{E}\right\rangle ^{2}=0$,
with pseudo-first-order rate $k_{EB_{i}}=k_{EB_{i}}^{\circ}\left[\textrm{S}_{i}\right]$,
where $k_{EB_{i}}^{\circ}$ is the rate constant for the binding of
substrate $\textrm{S}_{i}$.

\begin{figure}
\centering\includegraphics[width=3.25in]{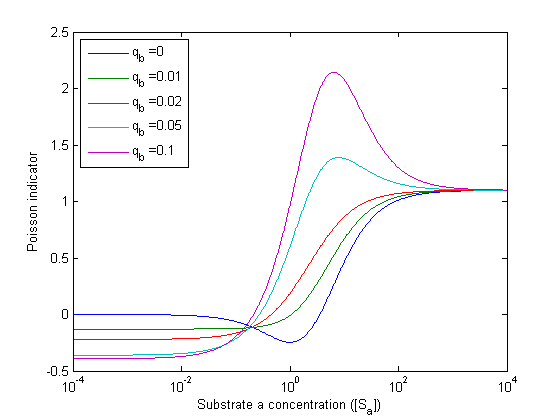}

\caption{\label{fig:MultSubstratesPlot}Plot of the Poisson indicator versus
$\left[\textrm{S}_{a}\right]$ for two competing substrates $\textrm{S}_{a}$
and $\textrm{S}_{b}$. The numerical parameters are $k_{EB_{a}}^{\circ}=1$,
$k_{EB_{b}}^{\circ}=1$, $q_{a}=0.5$, $\left[\textrm{S}_{b}\right]=1$,
$\left\langle \tau_{B_{a}}\right\rangle =0.55$, $\left\langle t_{B_{a}P}\right\rangle =1$,
$\left\langle t_{B_{b}E}\right\rangle =0.3$, $\left\langle t_{B_{b}P}\right\rangle =10$,
$\left\langle \tau_{B_{a}}^{2}\right\rangle =1.765$, $\left\langle t_{B_{b}E}^{2}\right\rangle =0.25$,
and $\left\langle t_{B_{b}P}^{2}\right\rangle =250$, with $q_{b}$
given in the legend. Note that the case of $q_{b}=0$ is equivalent
to substrate $\textrm{S}_{a}$ competing with inhibitor $\textrm{S}_{b}$.}
\end{figure}

The first moment for the overall waiting time in the presence of multiple
substrates is expressed as
\begin{equation}
\left\langle t\right\rangle =\frac{1+\sum_{i}k_{EB_{i}}^{\circ}\left[\mathrm{S}_{i}\right]\left\langle \tau_{B_{i}}\right\rangle }{\sum_{i}q_{i}k_{EB_{i}}^{\circ}\left[\textrm{S}_{i}\right]}
\end{equation}
where $\left\langle \tau_{B_{i}}\right\rangle =q_{i}\left\langle t_{B_{i}P}\right\rangle +\left(1-q_{i}\right)\left\langle t_{B_{i}E}\right\rangle $
is the first waiting time moment for the decay of bound/intermediate
state $\mathrm{B}_{i}$. Now, if we choose to examine the dependence
of the Poisson indicator on the concentration of a single substrate
$\left[\textrm{S}_{k}\right]$, it will have the following functional
form:
\begin{equation}
\mathcal{P}\left(\left[\mathrm{S}_{k}\right]\right)=\frac{\frac{A}{\left[\textrm{S}_{k}\right]^{2}}+\frac{B}{\left[\textrm{S}_{k}\right]}+C}{\left(\frac{1}{\left[\textrm{S}_{k}\right]}+D\right)^{2}}\label{eq:MultSubstrates_P}
\end{equation}
which notably departs from the functional form presented above for
the single-substrate and competitively inhibited cases (eq \ref{eq:FcnFormPoissInd}).
The constants $A$, $B$, $C$, and $D$, all independent of $\left[\textrm{S}_{k}\right]$,
have expressions
\begin{align}
A & =\aleph^{2}\left[-2\sum_{i\neq k}q_{i}k_{EB_{i}}^{\circ}\left[\mathrm{S}_{i}\right]\left\langle t_{B_{i}P}\right\rangle +\right.\nonumber \\
 & \,\,\,\,\,\,\,\,\,\,\,\,\,\,\,\,\,\left.\sum_{i\neq k,j\neq k}k_{EB_{i}}^{\circ}\left[\mathrm{S}_{i}\right]q_{j}k_{EB_{j}}^{\circ}\left[\mathrm{S}_{j}\right]\left(\left\langle \tau_{B_{i}}^{2}\right\rangle -2\left\langle \tau_{B_{i}}\right\rangle \left\langle t_{B_{j}P}\right\rangle \right)\right]\\
B & =\aleph^{2}\left[-2q_{k}k_{EB_{k}}^{\circ}\left\langle t_{B_{k}P}\right\rangle +\sum_{i\neq k}q_{i}k_{EB_{k}}^{\circ}k_{EB_{i}}^{\circ}\left[\mathrm{S}_{i}\right]\left(\left\langle \tau_{B_{k}}^{2}\right\rangle -2\left\langle \tau_{B_{k}}\right\rangle \left\langle t_{B_{i}P}\right\rangle \right)+\right.\nonumber \\
 & \,\,\,\,\,\,\,\,\,\,\,\,\,\,\,\,\,\left.\sum_{i\neq k}q_{k}k_{EB_{k}}^{\circ}k_{EB_{i}}^{\circ}\left[\mathrm{S}_{i}\right]\left(\left\langle \tau_{B_{i}}^{2}\right\rangle -2\left\langle \tau_{B_{i}}\right\rangle \left\langle t_{B_{k}P}\right\rangle \right)\right]\\
C & =\aleph^{2}q_{k}\left(k_{EB_{k}}^{\circ}\right)^{2}\left(\left\langle \tau_{B_{k}}^{2}\right\rangle -2\left\langle \tau_{B_{k}}\right\rangle \left\langle t_{B_{k}P}\right\rangle \right)\\
D & =\aleph k_{EB_{k}}^{\circ}\left\langle \tau_{B_{k}}\right\rangle 
\end{align}
where $\left\langle \tau_{B_{i}}^{2}\right\rangle =q_{i}\left\langle t_{B_{i}P}^{2}\right\rangle +\left(1-q_{i}\right)\left\langle t_{B_{i}E}^{2}\right\rangle $
is the second waiting time moment for the decay of bound/intermediate
state $\mathrm{B}_{i}$, and we have defined
\begin{equation}
\aleph=\frac{1}{1+\sum_{i\neq k}k_{EB_{i}}^{\circ}\left[\textrm{S}_{i}\right]\left\langle \tau_{B_{i}}\right\rangle }
\end{equation}
As a simple example of the behavior of the Poisson indicator in the
presence of multiple substrates, the Poisson indicator is calculated
for two competing substrates $\textrm{S}_{a}$ and $\textrm{S}_{b}$.
In Figure \ref{fig:MultSubstratesPlot}, the Poisson indicator is
plotted against the concentration of $\textrm{S}_{a}$ at a fixed
concentration of $\textrm{S}_{b}$. We note that in this plot, $\left\langle t_{B_{b}E}^{2}\right\rangle $
and $\left\langle t_{B_{b}P}^{2}\right\rangle $ are held fixed while
$q_{b}$ is varied, causing $\left\langle \tau_{B_{b}}^{2}\right\rangle $
to also vary, but $q_{b}$ could instead be varied while holding $\left\langle \tau_{B_{b}}^{2}\right\rangle $
fixed.

For $\left[\textrm{S}_{a}\right]=0$, the single-substrate result
at fixed $\left[\textrm{S}_{b}\right]$ is obtained as $\mathcal{P}\left(\left[\textrm{S}_{a}\right]=0\right)=A$,
which can be nonzero, differing from $\mathcal{P}\left(\left[\textrm{S}\right]=0\right)$
for the above two cases. In the limit of saturating $\left[\textrm{S}_{a}\right]$,
the single-substrate form for $\mathcal{P}_{\left[\textrm{S}_{a}\right]\rightarrow\infty}$
(eq \ref{eq:largeSlimit}) is obtained. It should be noted that earlier
results are recovered in the appropriate limits: setting $\left[\textrm{S}_{i\neq k}\right]=0$
recovers the single-substrate expression for $\mathcal{P}\left(\left[\textrm{S}_{k}\right]\right)$
(eq \ref{eq:GenChnRxnPoissonInd}). In addition, for only two competing
substrates $\textrm{S}_{a}$ and $\textrm{S}_{b}$, as in Figure \ref{fig:MultSubstratesPlot},
setting the branching probability $q_{b}=0$ recovers the competitive
inhibition result for $\mathcal{P}\left(\left[\textrm{S}_{a}\right]\right)$
(eqs \ref{eq:InhFcnlForm}-\ref{eq:Inh_C}), where $\left[\textrm{S}_{b}\right]$
corresponds to the inhibitor concentration. In fact, $\mathcal{P}\left(\left[\textrm{S}_{a}\right]\right)$
can achieve a local maximum similar to that shown in Figure \ref{fig:GenEnzInhPlot}(a)
for competitive inhibition. We identify the presence of such a maximum
for a renewal process as a signature of competitive binding, either
between a substrate and an inhibitor or between multiple substrates.
Finally, if the substrates are taken to be identical, that is $\left[\textrm{S}\right]=\left[\textrm{S}_{1}\right]=\left[\textrm{S}_{2}\right]=\ldots=\left[\textrm{S}_{n}\right]$,
then eq \ref{eq:MultSubstrates_P} describes the Poisson indicator
for an enzymatic reaction of a single substrate with multiple, parallel
pathways, nearly analogous to earlier results for ion channel statistics.\cite{Chaudhury2013}

\section{V. Concluding Remarks}

A general methodology for calculating second moments for the waiting
time between reaction events has been introduced and applied to the
analysis of enzymatic reactions. All of the flexibility conferred
by the self-consistent pathway analysis method (paper 1)\cite{Cao2008}
is retained, and the approach can be applied to many diverse cases.
Our approach is currently restricted to renewal processes but will
be extended to nonrenewal processes in a subsequent paper (paper 3).
In the current study, the principal results concern a generic enzymatic
reaction as well as the first explicit calculations (to our knowledge)
of higher-order waiting time moments for the more complex cases of
competitive inhibition and multiple substrates without assuming all
states undergo Poissonian decay. The use of a generic model of enzyme
catalysis allows the determination of the maximum information content
of measurements of the Poisson indicator and first waiting time moment.
Furthermore, analytical expressions for the Poisson indicator as a
function of substrate concentration allow connections to be made between
experimental data and kinetic models.

Our specific findings are summarized as follows: (i) based upon fitting
to the functional forms of the first two waiting time moments, the
non-Poissonian kinetics are generally underdetermined to second order
but can be specified under certain circumstances. (ii) For a generic
enzymatic scheme with an arbitrary intermediate topology, sub-Poissonian
statistics can always (for non-trivial kinetics) be achieved for a
certain range of substrate concentrations, even when branching occurs
out of the intermediate state(s). (iii) We have identified a generic
minimum of the Poisson indicator (with respect to substrate concentration),
and this can be used to tune substrate concentration to the stochastic
fluctuations, attaining optimal turnover statistics, and to estimate
the largest number of underlying consecutive rate steps in a turnover
cycle. (iv) At high and low substrate concentration, the Poisson indicator
reflects the effective reduction of the generic enzymatic scheme based
upon the rate-determining process. (v) We have identified a local
maximum of the Poisson indicator as a function of substrate concentration
for a renewal process as a signature of competitive binding, either
between a substrate and an inhibitor or between multiple substrates.
Our analysis may be easily extended to other single-molecule experiments,
offering the same benefits. In particular, application to the study
of motor proteins may be fruitful due to the presence of reaction
steps dependent upon substrate concentration as well as the applied
mechanical force.\cite{Keller2000}

\section*{Acknowledgements}

This work was supported by the Singapore-MIT Alliance for Research
and Technology (SMART) and the NSF (Grant No. CHE-1112825). D. E.
P. acknowledges support from the NSF Graduate Research Fellowship
Program.

\end{document}